\documentclass[manuscript]{aastex}

\slugcomment{Submitted to PASP}

\shorttitle{Characterization of Transiting SuperEarths}
\shortauthors{Deming et al.}

\begin{document}


\title{Discovery and Characterization of Transiting SuperEarths Using
an All-Sky Transit Survey and Follow-up by the $James~Webb~Space~Telescope$}


\author{D.~Deming\altaffilmark{1}, S.~Seager\altaffilmark{2}, J.~Winn\altaffilmark{3,4},
 E.~Miller-Ricci\altaffilmark{5}, M.~Clampin\altaffilmark{6}, D.~Lindler\altaffilmark{7},
 T.~Greene\altaffilmark{8}, D.~Charbonneau\altaffilmark{5}, G.~Laughlin,\altaffilmark{9},
 G.~Ricker\altaffilmark{4},
 D.~Latham\altaffilmark{5}, \& K.~Ennico\altaffilmark{8}}

\altaffiltext{1}{Solar System Exploration Division, Goddard Space Flight Center, Greenbelt, MD 20771}
\altaffiltext{2}{Department of Earth, Atmospheric \& Planetary Sciences, and Department of Physics, \\
  \indent~~~~Massachusetts Institute of Technology, 77 Massachusetts Ave, \\ 
   \indent~~~~Cambridge MA 02139-4307}
\altaffiltext{3}{Department of Physics, Massachusetts Institute of Technology, 77 Massachusetts Ave, \\ 
  \indent~~~~Cambridge MA 02139-4307}
\altaffiltext{4}{Kavli Institute for Astrophysics and Space Research, 
  Massachusetts Institute \\ \indent~~~~of Technology, 77 Massachusetts Ave, Cambridge MA 02139-4307}
\altaffiltext{5}{Harvard-Smithsonian Center for Astrophysics, Cambridge MA 02138}
\altaffiltext{6}{Exoplanet and Stellar Astrophysics Laboratory, Goddard Space Flight Center, \\ 
  \indent~~~~Greenbelt MD 20771}
\altaffiltext{7}{Sigma Scientific Corporation, Goddard Space Flight Center, Greenbelt MD 20771}
\altaffiltext{8}{Space Science Division, Ames Research Center, Moffett Field CA 94035-1000}
\altaffiltext{9}{UCO Lick Observatory, Santa Cruz CA 95064}


\begin{abstract}
Doppler and transit surveys are finding extrasolar planets of ever
smaller mass and radius, and are now sampling the domain of
superEarths ($1-3R_\oplus$). Recent results from the Doppler surveys
suggest that discovery of a transiting superEarth in the habitable
zone of a lower main sequence star may be possible.  We evaluate the
prospects for an all-sky transit survey targeted to the brightest
stars, that would find the most favorable cases for photometric and
spectroscopic characterization using the {\it James Webb Space
Telescope} (JWST). We use the proposed {\it Transiting Exoplanet
Survey Satellite} (TESS) as representative of an all-sky survey. We
couple the simulated TESS yield to a sensitivity model for the MIRI
and NIRSpec instruments on JWST.  Our sensitivity model includes all
currently known and anticipated sources of random and systematic error
for these instruments.  We focus on the TESS planets with radii
between Earth and Neptune.  Our simulations consider secondary eclipse
filter photometry using JWST/MIRI, comparing the $11-$ and $15\,\mu$m
bands to measure CO$_2$ absorption in superEarths, as well as
JWST/NIRSpec spectroscopy of water absorption from $1.7-$ to
$3.0\,\mu$m, and CO$_2$ absorption at $4.3\,\mu$m.  We find that JWST
will be capable of characterizing dozens of TESS superEarths with
temperatures above the habitable range, using both MIRI and NIRspec.
We project that TESS will discover about eight nearby habitable
transiting superEarths, all orbiting lower main sequence stars.  The
principal sources of uncertainty in the prospects for JWST
characterization of habitable superEarths are superEarth frequency and
the nature of superEarth atmospheres. Based on our estimates of these
uncertainties, we project that JWST will be able to measure the
temperature, and identify molecular absorptions (water, CO$_2$) in $1$
to $4$ nearby habitable TESS superEarths orbiting lower main sequence
stars.

\end{abstract}


\section{Introduction}

In the quest to measure the spectrum of a habitable exoplanet, the
observer must successfully disentangle the planetary flux from that of
the central star.  Two methods are currently in use for the direct
study of gas-giant exoplanets, and practitioners of each are working
to apply these methods to the study of rocky exoplanets.  In the first
technique, the planet and star are separated spatially through high
angular resolution imaging \citep{kalas, marois}.  This method favors
systems wherein the angular separation is as large as possible.  The
second method obviates the need for high angular resolution imaging by
studying the combined light of the planet and star in transiting
systems.  In these systems, the planet and star undergo periodic
mutual eclipses (see \citealp{charb07} for a review), and the emission
from each is subsequently separated with the knowledge of the
previously-characterized orbit.  The primary disadvantage of the
latter approach is obvious: only a small fraction of exoplanets will
have their orbits aligned so as to undergo eclipses as viewed from the
Earth.

The transit method offers several advantages that motivate its further
study.  The first is that of technological simplicity, since it does
not require the development of extreme-contrast ratio direct imaging.
The second is that of scientific impact: eclipsing systems, for which
the photometric transit and stellar radial velocity orbit have been
observed, permit a geometric determination of the planetary radius and
a dynamical estimate of the planetary mass by a means that is
virtually free of astrophysical assumption.  For such systems, the
interpretation of the hemisphere-averaged spectrum is likely to be
more scientifically fruitful than for systems lacking direct
measurements of the masses or radii.  For example, without a knowledge
of the planet radius, there exists a degeneracy between the emitting
area and surface flux (see e.g.,\,\citealp{kalas}).  More generally,
the bulk composition and physical structures of transiting exoplanets
are likely to be well constrained, and inferences about the atmosphere
(for example, its chemical composition) are likely to be far more
penetrating than for cases in which only the spectrum is available.

Over the past 7 years, the characterization of the atmospheres of
Gyr-old planets orbiting solar peers has proceeded rapidly and
exclusively through the study of combined light of the planet and star
in transiting systems.  A partial list of such successful studies
includes \citet{charb02, charb05, charb08}, \citet{deming05, deming06,
deming07}, \citet{grillmair07, grillmair08}, \citet{harrington06,
harrington07}, \citet{knutson07, knutson08, knutson09},
\citet{richardson}, \citet{swain08, swain09}, \citet{tinetti}, and
\citet{vidal-madjar}.  All of the detections listed in the previous
sentence were accomplished with either the {\it Hubble Space
Telescope} or {\it Spitzer Space Telescope}, two general purpose,
space-based facilities, neither of which were optimized for exoplanet
studies (indeed, these studies were not anticipated during the
development of either facility).

Since there is currently intense interest in rocky superEarth
exoplanets,\footnote{We define a superEarth to be a rocky or icy planet
having a radius less than $3R_\oplus$} it is desirable to explore the
factors that will limit their discovery and study using
transit/eclipse methodology. We consider the coming {\it James Webb
Space Telescope} (JWST) and ask what it could accomplish for studies
of the atmospheres of transiting exoplanets, especially for smaller,
rocky, and potentially habitable worlds.  We note that neither {\it
Hubble} nor {\it Spitzer} served as the discovery observatory for the
planets they characterized: rather, the tasks of discovery and
characterization were split, with ground-based surveys pursuing the
former and these two spacecraft conducting the latter.  Similarly, we
anticipate that JWST will be able to undertake characterization
studies of the atmospheres of habitable worlds, but only if these
objects are identified in nearby, transiting systems by a separate
discovery effort. Since the most favorable systems are the closest
ones, and very high photometric precision is needed for long time
periods, a bright star, all-sky, space-borne survey such as the
proposed {\it Transiting Exoplanet Survey Satellite (TESS)} is the
natural starting point for our study.

The specific purpose of this paper is to evaluate the potential for
discovery and characterization of exoplanets by using an all-sky
transit survey and JWST follow-up.  Our simulation includes planets
ranging in size from Earth to Jupiter. JWST follow-up of Jupiter-size
transiting planets would be of enormous scientific
utility. Nevertheless, we here focus on planets of Neptune size and
smaller. Our analysis includes `hot superEarths' as well as those in
the habitable zone (HZ)\footnote{We use the term habitable to be
synonymous with a thermal equilibrium temperature range of 273 to 373
Kelvins, with no other requirements.}. We evaluate whether the sources
of uncertainty for superEarth discovery and characterization are
primarily astrophysical (e.g., occurrence rate for superEarths), or
technological (sensitivity of instrumentation and surveys).  Our
evaluation begins by projecting the yield of TESS.  Although TESS is
still at the proposal stage, it provides the most well-studied basis
from which to project the yield of a space-borne, bright star,
all-sky, transit survey.  We couple the TESS planets to sensitivity
calculations for specific observing modes of the Near-Infrared
Spectrograph (NIRSpec) and Mid-Infrared Instrument (MIRI) instruments
on JWST.  Our intent is $not$ to define and compare the relative
merits of all potential modes of JWST exoplanet observations.
Instead, we concentrate on the likely success for exoplanet
characterization using two of the most obvious observational modes.

Potential JWST exoplanet transit observations have been discussed by
several authors \citep{clampin09, greene_1, kaltenegger, lunine,
seager08, valenti05, valenti06, valenti08}. Our treatment here goes
beyond previous work in that we specifically include simulations of an
all-sky survey, coupled to a characterization simulation that includes
the JWST's pointing constraints (field-of-regard), and anticipated
JWST instrument- and detector-related systematic errors, based on
experience with {\it Spitzer} and current engineering data from the
JWST instruments.

Sec.~2 presents an estimate of the likely distance to the closest
transiting habitable exoplanet, and Sec.~3 describes our statistical
model for the yield of TESS. Sec.~4 discusses a simple model for
exoplanet characterization using transits and eclipses, and later
sections develop more exhaustive simulations of JWST characterization
of TESS planets: Sec. 5 gives our sensitivity and noise models for the
JWST instruments, and Sec.~6 applies these noise models to the TESS
planets. Sec.~7 gives our conclusions and comparisons to other work.

\section{The Closest Transiting Habitable Exoplanet}

It is of interest to consider the probable distance of the closest
habitable planet that transits, assuming that we discover $all$ nearby
transiting planet systems. The probability of a transiting habitable
planet within distance $d$ can be calculated from the space density of
nearby stars, if we know the distribution of planets {\it vs.} orbital
size for each stellar spectral type.  Because these distributions are
not securely known, we adopt a very general assumption.  We assume
that a fraction $f$ of stars host exactly one planet in their HZ, and
we place half of them at the inner HZ boundary, and half at the outer
HZ boundary.

The probability of at least one transiting planet within a sphere of
radius $d$ centered on the Sun equals $1-\prod_i{p_i}$, where $p_i$ is
the probability that $no$ planet transits star $i$, and $i$ indexes
the stars within distance $d$.  Given the above assumptions on the
orbital radii of the planets, and the transit probability for each
planet ($R_{star}/a$), this is a straightforward calculation. We adopt
circular orbits, and the stellar space densities versus spectral type
described in Sec.~3.  The results are shown in Figure~1 for planet
frequencies of $f=1.0,~0.3$ and $0.1$. This calculation shows that if
every star hosts a planet in the HZ, then the closest transiting
habitable planet is likely (i.e., probability 0.5) to be found at
$d=5$\,pc.  This distance increases to $d=7$ and $10$ pc for $f=0.3$
and $0.1$.  Moreover, even for a HZ frequency as low as $f=0.01$ per
star, the probable distance for the nearest example (not illustrated)
is $d=22$ pc. These calculations suggest good prospects for JWST
characterization even if HZ planets are moderately scarce.  However,
there remains the practical problem of whether a given transit survey
such as TESS can necessarily find the nearest examples.

\section{Yield of an All-Sky Transit Survey}

We here simulate the yield of an all-sky survey targeted primarily to
bright ($V<13.5$) stars, and we elaborate on how the superEarth yield
of such a survey depends on superEarth occurrence and orbital
distribution properties.  We use the TESS survey \citep{ricker} as a
practical example of an all-sky bright star survey capable of
detecting transiting superEarths. TESS is proposed to view the sky
from near-Earth equatorial orbit, using an array of 6 wide-field ($18
\times 18$ degree) refractive CCD cameras on a single spacecraft. TESS
monitors $2.5 \times 10^6$ bright ($V < 13.5$) stars, with special
attention to lower main sequence stars. During each orbit, TESS will
monitor a 72-degree strip of right ascension, using several pointings.
Each star in the TESS catalog will be monitored for at least 72 days.
Since TESS requires two transits to identify a planet, it is
insensitive to orbital periods exceeding 72 days.

\subsection{Making Transiting Planets}

Our Monte-Carlo model of the planets to be found by TESS considers main sequence 
host stars from spectral types F5 through M9. We adopt the absolute
visual magnitude {\it vs.} spectral type relation from \citet{henry},
and the main sequence luminosity function from \citet{reidhawley}.  We
convert spectral type to $T_{eff}$, stellar mass and radius, using
Table 4.1 of \citet{reidhawley}.

We make planets by first generating a set of Monte-Carlo stars in a
cubical volume of $2000^3$\,pc$^3$ centered on the Sun. This cube is
larger than the TESS search space and it contains all planets that TESS
will find.  We assign galactic coordinates (X, Y, Z) to each star by
placing them randomly within this cube, constrained by their space
densities at Z=0 {\it vs.}  spectral type, and enforcing an
exponential distribution in height above the galactic plane (Z), with
200 pc scale height. Since transiting planets require mass
measurements via high precision radial velocities, we eliminate the
visually faintest stars (having $V > 13.5$), but we keep all stars at
distances closer than 35 pc \citep{charbdem}, {\it regardless of their
visual magnitude}.  Planet-hosting M-dwarfs can lie at close distances
and still be visually faint, because their spectral energy
distribution peaks in the IR.  We reason that superEarths transiting
nearby M-dwarfs will be so important that their radial
velocities will be measured by some means, regardless of their visual
magnitude.  For example, IR spectroscopy could be used when it is
sufficiently developed to achieve the precision needed for superEarths
\citep{blake}.

We assign exactly one planet to each star.  We use various simple
distribution formulae for orbital size and planetary radius, because
the actual distributions are poorly known.  An important rationale for
our work is to define the expected yield of superEarths under various
orbital size distributions.  Our default distribution in orbital size
uses a uniform probability density in $\log(a^{'})$, where
$a^{'}=a\,(L_{\odot}/L_*)^{1/2}$, between $0$ and $-1.3$ in
$\log(a^{'})$.  One rationale for the stellar luminosity scaling
factor is that it produces a more compact planet distribution orbiting
low-mass stars, consistent with evidence for close-in planet
formation in some M-dwarf planetary systems \citep{forbrich}. Also, this
scaling puts the same number of planets in each temperature zone, and
is a natural way to cast the problem when interested in the HZ. At the
opposite extreme, we also explore the effect of a distribution whose
probability density for orbital size is uniform in $a$.

Because the magnitude of the transit signal depends on the radius of
the planet, we use radius as the independent variable in our planet
distributions.  We distribute planet radii ($R_p$) with a uniform
probability density in $\log(R_p/R_\oplus)$, between limits of $0$ and
$1.1$ in the log (these limits correspond to Earth and Jupiter-sized
planets, respectively). Surface gravity is needed for subsequent
simulations (Secs. 5 \& 6).  To that end, we calculate the mass from
the radius by assigning a bulk composition, and inverting the
mass-radius relations of \citet{seager07}.  When $R > 2R_\oplus$, we
use an icy bulk composition (ocean planets), and when $R > 3R_\oplus$
we add a H-He envelope having a depth of $1R_\oplus$.  Planets having
radii $R_p < 2R_\oplus$ are assigned bulk compositions of either
Earth-like (silicate), or Mercury-like, with equal probability. Using
this radius distribution, approximately $40$\% of the simulated
planets are superEarths (defined by TESS as having $R \leqq
3R_\oplus$).

For each planet, we calculate whether it transits based on random
selection in proportion to the transit probability. Transit
probability is the ratio of the stellar radius to the planet's orbital
radius, and we used circular orbits for all planets. Using circular
orbits is conservative, because transit probability increases for
eccentric orbits \citep{barnes}. We also assign a random impact
parameter to each transit, uniformly distributed between zero and
unity. From the impact parameter, we calculate the duration of each
planet's transit. Finally, we calculate the temperature of each
planet, based on its orbit radius and stellar luminosity, scaling from
a 287K value at 1\,AU with solar luminosity.

\subsection{Finding Transiting Planets}

In deciding whether TESS will discover a particular transiting planet,
we take the duty cycle and sensitivity of the mission into account.
TESS will observe a given location in the sky once per 96-minutes, for
72 days.  For each transiting planet, we calculate the in-transit
times from the planet's orbital period, assigning a random phase.  We
then tabulate the number of transits that overlap the TESS observing
cadence, using a 10-minute time resolution. Those planets seen by TESS
during at least two transits are counted as detected, providing they
meet the SNR requirement.

To calculate the observed SNR of each transit, we used Phoenix model
atmospheres \citep{hauschildt} for stars of each spectral type, and
derived the flux incident at the TESS telescopes. The flux into the
$84.6$\,cm$^2$ effective collecting area of the TESS telescopes
(including optics throughput and detector quantum efficiency) was
integrated over the 600-1000 nm bandpass of the TESS detectors, to
establish a count rate for each transiting system. Given the duration
of the transit, we calculate the photon-limited SNR for the aggregate
of the phased transits, and we require SNR for the transit depth to be
at least $7$. We also eliminate planets having transit depths less
than TESS's estimated systematic noise floor (100 ppm).

It is of interest to determine whether a mission in near-Earth orbit
like TESS has an adequate sampling cadence to find transits, or
whether it misses many.  In our simulation, only a very small fraction
of planets ($<< 0.1$\%) exhibiting more than two transits within the
72-day TESS observing period were missed because of sampling. Those
few cases include grazing transits of short period planets orbiting small
stars, where the transit duration was brief (typically 30 minutes).
With short transit durations, a small fraction can be resonantly out
of phase with the TESS 96-minute sampling interval, but the number of
such cases is negligible.

We used the above simulation procedures to produce a Monte-Carlo
realization of the TESS yield, adopting a frequency of one planet
per star.  The real planet frequency will differ from unity, but our
results are linearly scalable to other values.

\subsection{Distribution of Planets and Implications}

Figure~2 shows the number of planets that TESS should find, versus
their radii ($R_p$), under two different assumptions concerning their
orbital sizes. Using our default distribution of orbit sizes,
scaling with $a^{'}$ (see above), TESS finds 16,244
planets.  The radii of the detected planets peak just above the
exoNeptune regime, at $6\,R_\oplus$.  We believe that our default
distribution for $a$ is a resonable estimate. But using a
distribution of orbital size that is uniform in $a$, the number of
detected planets drops to 6845, a factor of 2.4 less.  The decrease is
relatively independent of planet radius, and occurs because a uniform
distribution in $a$ overweights larger orbital sizes compared to a
uniform distribution in $\log a^{'}$. Larger orbital sizes translate to
smaller transit probabilities, and thus the yield of the survey is
reduced.

The total number of small planets detected is greatly affected by the
photometric sensitivity of the survey. Our default radius distribution
has uniform probability density in the logarithm of $R_p$, and that
distribution (not illustrated) slopes downward on Figure~2, with more
planets occurring in bins at small radius.  But many of those small
planets orbit relatively distant faint stars, and are not detected by
TESS because their transits are not measured to the requisite SNR. For
example, our default distributions produce 7857 transiting superEarths
with radii between 2 and 3\,$R_\oplus$.  TESS's total for the 2- to
3\,$R_\oplus$ bin is 409 out of the 7857 due to incompleteness at the
largest distances. (TESS is flux-limited, and is not designed to be a
volume-limited survey.)  The total yield of superEarths from an
all-sky transit survey is extremely sensitive to the light grasp and
photometric precision of the survey, to a greater extent than
uncertainties involving the superEarth orbital size distribution. For
gas-giant planets, SNR is not a limiting factor.  Instead, the number
of transiting giant planets detected is sensitive to their orbital
architecture, and to galactic structure.

Finally, we evaluated the effect of visual {\it vs.} IR brightness for
M-dwarfs. Our default simulation of TESS finds 320 superEarths within
35 pc of the Sun, regardless of their star's visual magnitude. If we
strictly require $V < 13.5$ on the grounds that radial velocity
confirmation would have to be done in the visible spectral range, then
the number of TESS superEarths within 35 pc drops to 90.  This implies
that TESS is efficient for M-dwarf transits due to its bandpass extending
to 1000 nm.  It also implies that successful development of precision
IR radial velocity techniques may significantly increase the number of
confirmed superEarths from a transit survey of bright stars.

\subsection{Completeness of TESS for Nearby Transiting Habitable SuperEarths}

The potential discovery of habitable superEarths from the TESS survey
is of special interest. We calculated the completeness of TESS for
superEarths ($1-3R_\oplus$) by running our Monte-Carlo simulation 500
times, and tabulating the fraction of habitable superEarths that TESS
finds versus stellar mass and distance. The top panel of Figure~3
shows the completeness of TESS for transiting habitable superEarths
{\it vs.} stellar spectral type, in a volume out to 35 pc.  All of the
habitable superEarths found by TESS orbit M-dwarfs.  The completeness
of TESS for $M > 0.5M_\odot$ drops steeply because the habitable zone
moves to longer orbital periods, where TESS does not sample two
transits.  Although TESS misses habitable superEarths orbiting stars
more massive than about $0.5M_\odot$, it efficiently finds the
transiting habitable superEarths nearest to our Sun. The completeness
of TESS versus distance, including stars of all spectral types, is
shown in the lower panel of Figure~3. TESS completeness is $93\%$ at
10 pc, $80\%$ at 20 pc, and $62\%$ at 35 pc.  It falls steeply at
greater distances. The high completeness at near distances is
consistent with the fact that the nearest stars are predominately
M-dwarfs. TESS misses the few transiting habitable superEarths that orbit
stars more massive than about $0.5M_\odot$. These misses are few in
number because such stars have lower space densities than M-dwarfs,
and planets in their habitable zones have lower probabilites to
transit compared to M-dwarf habitable planets.

\section{Overview of SuperEarth Characterization via Transits}

The characterization of transiting exoplanets relies primarily on
observations at transit (planet partially eclipsing star) as well as
secondary eclipse (star eclipsing planet). Besides the mass and radius
determined from transit photometry (e.g., \citealp{charb06, winn}),
the atmosphere can be characterized via its transmission spectrum
during transit \citep{charb02, redfield, tinetti, swain08} and the
emergent atmospheric spectrum can be measured using photometry and
spectroscopy at secondary eclipse \citep{charb05, charb08, deming05,
deming06, harrington07, knutson08, richardson, grillmair07,
grillmair08}. To estimate the sensitivity limits for these techniques
as applied to superEarths, it is useful to first consider a simple
characterization model.  In this initial model, we assume that both
the planet and star radiate as blackbodies. Some exoplanet atmospheric
models predict that the atmospheric annulus of a transiting exoplanet
will be opaque in the strongest lines over a height range of $\sim 5H$
\citep{seagersasselov,miller-ricci}, where $H$ is the pressure scale
height.  Hence we approximate the magnitude of transit absorption,
relative to a transparent continuum, as being equal to the area of an
annulus of height $=5H$. If the atmosphere is opaque over fewer scale
heights, then the SNR in this simple model would be reduced in direct
proportion.  We calculate $H$ using a temperature of 323K, and a mean
molecular weight of 22.  These parameters are in an intermediate range
for superEarths: the temperature is mid-way between the freezing and
boiling point of water, and the atmospheric mean molecular weight is
intermediate between free-hydrogen-rich and free-hydrogen-depleted
(e.g., pure carbon dioxide) atmospheres \citep{miller-ricci}.  Both
this initial calculation, as well as our more detailed simulations
(Secs. 5 \& 6), adopt cloud-free atmospheres.  We here use a
superEarth mass of $10\,M_\oplus$, and a `dirty ocean planet' bulk
composition, intermediate between rocky and icy. Using the mass-radius
relations given by \citet{seager07} yields a planet radius of
$2.3\,R_\oplus$.  The depth of secondary eclipse is approximated as
the planet-to-star area ratio, times the ratio of Planck functions.

In analogy with Hubble Deep Field investigations \citep{gilliland},
JWST characterization of a habitable superEarth would justify a
large allocation of observing time, covering many transits and
eclipses. We here adopt a total observation time of 200 hours.
Calculation of signal-to-noise ratio (SNR) for each transit and
eclipse observation requires accounting for the uncertainty of the
out-of-transit (or eclipse) baseline flux level. However, for the most
interesting cases (i.e., habitable planets) the orbit periods are
relatively long compared to hotter planets. Habitable planet
observations will be limited by the number of transits that are
visible to JWST within its 5-year mission, not limited primarily by
total observing time in hours.  We therefore calculate the SNR using
the condition that the 200~hours applies only to the
in-transit/eclipse period, and that additional time will be available
to measure baselines to a precision that does not significantly
increase the error of the results.

Since we are here concerned with relatively nearby stars observed with
a large-aperture space telescope, the observations are in the
high-flux limit and we expect the dominant random noise source to be the
photon noise of the host star and thermal background radiation, and we
include these noise sources for this zeroth order model in the same
detail as for our full MIRI and NIRSpec noise models (Secs. 5 \& 6).  We adopt a
telescope collecting area of $25$\,m$^{2}$, and an end-to-end efficiency
(electrons out/photons in) of $0.3$ (typical for {\it Spitzer} and expected for JWST).

Figure~4 shows the results from this zeroth order calculation, as
contours of constant SNR versus the distance to the system and the
temperature (hence, mass and radius) of the host star. Transmission
spectroscopy is more favorable at shorter wavelengths, because the
transmission signal is proportional to the intensity of the stellar
disk.  We used water absorption at 2\,$\mu$m for the top panel of
Figure~4. We summed the FWHM of several strong bands in model
transmission spectra \citep{miller-ricci} as an estimate of the number
of wavelength points ($160$) that would be optically thick over $5H$
in height.  This signal is detectable to SNR\,$\sim\,10$ for systems
out to $\sim$\,12 pc, for a wide range of stellar temperatures.
Moreover, the SNR contours bulge outward to higher SNR for the lower
main sequence, showing the `small star effect' \citep{charbdem}.  The
lower panel of Figure~4 gives the SNR for secondary eclipse photometry
at 15\,$\mu$m, near the peak of the planet's thermal
emission. Detection of the secondary eclipse to SNR\,$\sim$\,10
requires a distance closer than $10$ pc for hotter stars, but the
small-star effect extends $SNR\,\sim\,10$ to $\sim\,30$ pc for cooler
stars like M-dwarfs.

Figure~5 shows the results from our zeroth order model in the case
where we raise the planet's temperature to $T=500$K, increase the mass
to $20\,M_\oplus$, and lower the mean molecular weight of the
atmosphere to 2.3 (hydrogen-helium composition, with oxygen at solar
abundance). This planet is a `hot superEarth', and the SNR for its
characterization is considerably more favorable than for habitable
superEarths.

We have overplotted some examples of TESS planets on Figure~4,
extracting those planets that closely bracket the parameters assumed
in generating the SNR contours.  For Figure~4, we overplot the TESS
planets having $R=2.3\pm 1.0\,R_{\oplus}$, and $T=323\pm50\,K$, i.e.,
habitable superEarths. For Figure~4, TESS finds $5$ habitable
superEarths that could be characterized by NIRSpec to SNR $\sim\,10$
or greater. A similar number of habitable superEarths could be
characterized by MIRI secondary eclipse photometry.  For Figure~5,
TESS finds many more hot ($T\ge 500$K) superEarths, and we have
overplotted examples of these.  As we will see in later Sections,
these numbers are in reasonable agreement with much more exhaustive
calculations.

\section{Sensitivity and Noise Models for JWST Instruments}

JWST is well suited to transiting planet characterization
\citep{clampin09}.  It will orbit at the L2 Lagrangian point, where
continuous observations will be possible without significant blocking
by the Earth. The telescope will experience a favorable thermal
environment, with low background emission.  Engineering studies have
defined the pixel response functions, and have developed a stringent
error budget for pointing jitter.  Moreover, JWST instruments will
incorporate direct-to-digital detector readouts, with minimal
suceptibility to electrical interference.  We have utilized JWST
Preliminary Design Review engineering estimates of telescope
performance, to couple our TESS yield calculations to simulations of
JWST superEarth characterization.

Our simulations specify that JWST will observe all transits and
eclipses of a given planet that are possible in principle during its
5-year mission \citep{gardner, clampin08}. The rationale is that the
maximum number can be objectively calculated, and readers can scale
the $SNR$ for smaller programs as they deem appropriate.  The maximum
number of transits/eclipses is particularly large for the hotter
planets, and scaling will show they can be observed to good SNR with
far below the maximum number of transits.

Since the galactic coordinates of each Monte-Carlo planetary system
are tagged by our simulation, we transform to ecliptic coordinates and
thereby calculate the time that each individual system is available in
the JWST field-of-regard. We simulate each of the observable
transits/eclipses, and we specifically include non-white noise sources
during observations of each transit/eclipse.  However, we expect that
the aggegate SNR for observations of multiple events will be
proportional to the square root of the number observed, because each
event is a relative measurement (in-transit/eclipse compared to
out-of-transit/eclipse), and the various transits/eclipses are
independent.  Hence our results can be adjusted to less complete
observational programs by scaling the aggregate SNR in proportion to
the inverse square-root of the number of events that are actually
observed.

In order to accurately evaluate the limits of characterization by
JWST, we must construct a realistic noise model for the observations.
In so doing, we draw on experience of {\it Spitzer} exoplanet
programs, and current engineering error budgets and data for the JWST
instruments, to simulate some effects that are forseeable for each
JWST instrument.  We concentrate on two of the many possible modes of
JWST observation, namely filter photometry at secondary eclipse using
MIRI, and transmission spectroscopy at transit using NIRSpec.

\subsection{MIRI}

JWST's Mid-Infared Instrument (MIRI, \citealp{wright}) will be the
primary resource for exoplanet secondary eclipses, since it covers the
$5-$ to $28\,\mu$m wavelength range where secondary eclipse contrast
is maximized. MIRI will provide filter imaging, low resolution grism
spectroscopy ($R\,\sim\,100$), and medium resolution spectroscopy
($R\,\sim\,2000$). We have concentrated our noise model on imaging
filter photometry, rather than MIRI spectroscopy, because we want to
explore a photometric technique in addition to spectroscopic
observations. This choice is conservative, because recording all
wavelengths simultaneously could in principle give MIRI spectroscopy a
factor of two improvement over photometric characterization.
				   
Our MIRI noise model includes thermal background emission from the
instrument, telescope, and sun shade, as calculated by
\citet{swinyard}, as well as zodiacal thermal emission from our own
solar system using the best available model \citep{kelsall} for the
dependence on ecliptic latitude. We adopt a detector Fowler-8 read
noise of 20 electrons per pixel, and we include $0.03$
electrons/sec/pixel dark current.  We use a total reflectivity for the
telescope optics of $0.88$, and a transmittance of $0.52$ for the MIRI
instrument optics including the filters.  The transmission curves of
the MIRI filters are not yet available, but we include the requirement
on their peak transmittance ($>0.75$) in the optics throughput, and we
use square bandpass functions matching their required FWHM.  We use
detector quantum efficiencies of $0.5$ and $0.6$ at 11 and 15\,$\mu$m,
respectively \citep{swinyard}.

Our JWST noise model also includes an anticipated source of systematic
error, based on {\it Spitzer} experience.  The Si:As detectors that
will be used in JWST/MIRI are similar to those in {\it
Spitzer}/IRAC, and JWST will have pointing jitter like {\it
Spitzer}, but projected to be of smaller amplitude.  Hence,
high-precision {\it Spitzer} photometry at 8\,$\mu$m provides our best
current proxy for systematic errors in JWST/MIRI photometry.

{\it Spitzer} exhibits a periodic oscillation in pointing, with a
1-hour period and tens of milli-arcsec amplitude. This pointing error
dithers the stellar image with respect to the detector pixels.  Since
the relative response of each pixel is imperfectly known (i.e., there
are inaccuracies in the flat-fielding), the pointing oscillation
causes an intensity oscillation in aperture photometry. Assuming
(conservatively) that the MIRI flat-fielding is not improved over {\it
Spitzer}, we can use the {\it Spitzer} data to calculate the magnitude
of this systematic error for MIRI photometry.  This requires modeling
some recent $Spitzer$ data before turning our full attention to
JWST/MIRI.

We developed a numerical model that simulates this effect.  For the
{\it Spitzer} data, we calculate the Fourier power spectrum of the
\citet{seagerdeming} data, shown in Figure~6.  This spectrum shows a
peak at a period of slightly more than 1~hour, close to the known
period of the telescope oscillation.  The power in this peak is
$1.1\%$ of the power integrated over all frequencies, hence the
amplitude of the 1-hour oscillation in intensity at $8\,\mu$m is
${{0.011}^{1/2}}\sigma=0.105\sigma$, where $\sigma$ is the standard
deviation of the intensity time series.  We know the amplitude of
pointing oscillations at this frequency, from measuring the stellar
image displacement in the \citet{seagerdeming} data.  The link between
pointing oscillation and intensity oscillation is the imperfect
pixel-to-pixel flat-fielding. Hence we have sufficient data to
determine the magnitude of pixel-to-pixel flat-fielding error for {\it
Sptizer}/IRAC. 

Our numerical model resamples {\it Spitzer's} $8\,\mu$m PSF to 10
times finer spacing, and convolves it with a simulated detector grid,
also resampled to 10 times finer spacing.  We do the convolution at a
series of pointing values, simulating the pointing oscillation in
synthetic time series photometry. The model includes flat-fielding
errors on the scale of the actual IRAC pixels, with the error per
pixel assigned from Gaussian random errors of a specified amplitude.
We vary that amplitude until the standard deviation of the simulated
time series photometry matches the observed amplitude in intensity
inferred from the Fourier analysis described above.  On this basis, we
estimate the IRAC flat-fielding error at $8\,\mu$m to be $0.4\%$.  We
use that value in our MIRI noise model.

Our MIRI noise model adopts a PSF from a ray-traced optical model of
the telescope plus instrument, for a central field location
(S. Ronayette, private communication). This PSF is polychromatic,
i.e., it is based on multiple wavelength samples over the $12.8\,\mu$m
filter bandpass. We spatially stretch it to represent the PSF at other
wavelengths. The properties of the JWST pointing jitter can be
anticipated from the engineering requirements \citep{osta}.  The
telescope body pointing is controlled at a 16\,Hz sampling rate, with
a 0.02\,Hz bandwidth.  The fine guidance sensor is also sampled at
16\,Hz, with a 0.6\,Hz bandwidth.  The $1\sigma$ amplitude of pointing
jitter above the control bandwidth is 4.2 milli-arcsec (mas) per axis,
and within the bandwidth it is 5.1\,mas per axis.  Uncontrolled drift
is specified to be less than 2.2\,mas per axis in a 10,000-second
science exposure.  We model the jitter within and above the control
bandwidth as Gaussian random error of the specified magnitude, and we
model the uncontrolled drift as monotonic and linear in time (worst
case). This produces a $1\sigma$ deviation of 7\,mas per axis for a
10,000-second exposure. We note that exoplanet observers find a
monotonic drift of much larger amplitude ($\sim$ hundreds of mas) in
{\it Spitzer} pointing over $\sim$\,tens of hours.

We use the above attitude control properties to generate pointing
errors for a long time series of synthetic MIRI photometry.  We adopt
highly oversampled (100x) realizations of the MIRI detector grid and
the PSF. The detector grid has $0.4\%$ uncompensated response
variation per original pixel. We shift the PSF relative to the
detector grid, in accord with the synthetic pointing errors, multiply
by the pixel sensitivities, and sum spatially to produce synthetic
aperture photometry that contains only this error source. We model
this error using a 10-second time resolution, but the actual exposure
time for bright stars may be shorter than 10-seconds.  Our noise model
implicitly assumes that short exposures can be co-added (on the
ground) to 10-second resolution, with no loss except for the increased
overhead entailed by frequent detector reading (see below).

Noise from a representative 6-hour set of 10-second exposures, is shown
in Figure~7.  We generated a 200-hour times series using this
methodology, and we draw a sequential portion of that noise reservoir,
starting at a random time, when modeling each eclipse for each
planet. Figure~7 includes the histogram of modeled intensity
fluctuations.  This histogram is distinctly non-Gaussian, but the
amplitude of this error source is generally small ($< 10^{-4}$).
Although JWST has a higher spatial resolving power than {\it
Spitzer} (tending to increase this error), it also has much finer
pixel scale, and much better telescope pointing control, which greatly
reduces this source of photometry error.

Some planets will orbit bright stars, filling the MIRI detector in
less than a 10-second exposure time.  These systems require increased
overhead to read-out the detector frequently in subarray mode.  We
calculate the exposure time required for each planet-hosting star to
fill the brightest pixel to $10^5$ electrons \citep{wright}.  Our
noise model includes a 50 msec overhead per subarray read, and thereby
accounts for the observing efficiency.

\subsection{NIRSpec}

The principal use of NIRSpec for transiting exoplanets will be to
measure the transmission spectrum of their atmospheres during
transit. We have modeled observations of transit water absorption in
the $1.6-$ to $3\,\mu$m region, and CO$_2$ absorption near
$4.3\,\mu$m, as observed using slit spectroscopy at a spectral
resolving power $R=1000$.  The optical design of NIRSpec has been
discussed by \citet{kohler}.  Our model adopts a total optical
transmission for the NIRSpec optics after the slit of $0.4$, and we
also include the wavelength-dependent grating blaze function.  As for
optical losses at the entrance slit, we note the recent plan to
include a $1.6 \times 1.6$ arc-sec entrance slit in NIRSpec, in
specific response to the potential for exoplanet spectroscopy.  No
slit losses are included in our noise model because this large
aperture will encompass virtually all of the energy in the telescope
PSF.

Our noise model for NIRSpec includes the limitation on exposure time
due to rapidly filling the detector wells (full well = 60,000
electrons).  We use the Phoenix model atmospheres to calculate the
time to achieve that signal level for the brightest pixel.  As in the
MIRI case, we model the systematic errors assuming a 10-sec time
resolution. Our model accounts for the number of times the detector
must be read in a 10-second exposure, and we include the time to read
a $16 \times 4096$-pixel subarray (0.85 seconds), since it reduces the
observing time efficiency and increases the impact of read noise.

The noise properties of the NIRSpec detectors have been discussed by
\citet{rauscher}. Our noise model for NIRSpec adopts a quantum
efficiency of 0.8 for these HgCdTe detectors, independent of
wavelength.  We include detector read noise (6 electrons per Fowler-8)
and dark current ($0.03$ e/sec) in our model, but these sources of noise
are not significant in comparison to source photon noise and
systematic error due to intra-pixel sensitivity variations. Recent
measurements of the intra-pixel sensitivity variations in these
detectors \citep{hardya} show that the detector pixels have maximum response near
pixel center, similar to the effect seen in the shortest wavelength
channels of $Spitzer$/IRAC \citep{morales-calderon}.  Figure~8 shows
this intra-pixel variation at a resolution of 0.1-pixels in each axis,
as reconstructed by one of us (D.L.) based on engineering measurements
at several locations in several pixels of the flight detectors.  Our
noise model synthesizes a large array of similar pixels, but we do not
have complete information on possible pixel-to-pixel variation in the
intrapixel curve.  We therefore make the reasonable assumption that
the average center-to-edge response variation is the same for all
pixels, but that variations around a `smooth' intrapixel curve will
differ from pixel to pixel.  We fit a quadratic to our available
center-to-edge data to define the smooth intrapixel curve (i.e., the
average pixel).  We measure the magnitude of deviations from this
smooth curve, and we add Gaussian random noise having that standard
deviation to the smooth curve, and thereby define the variations in
the response curve of each modeled pixel, at a spatial resolution of
0.1-pixels in each axis.

Because there will be pointing jitter in the telescope, the spectrum
will be dithered over the grid of detector pixels.  This will result
in an intensity variation at each wavelength, similar to the effect
seen in $Spitzer$/IRAC photometry at $3.6-$ and $4.5\,\mu$m
\citep{charb08}.  We model this process by using an optical model of
the NIRSpec PSF at $2\,\mu$m, and we scale the width of this PSF and
convolve it {\it at each wavelength} with our synthetic pixel grid,
using a spectrum of telescope pointing jitter and drift as described
above.  The optical model of the spectrograph PSF uses the telescope
PSF, and disperses it using a Gaussian kernel with a FWHM equal to the
spectral resolution. We resample both the PSF and the detector pixel
grid to $0.02-$pixel resolution, for maximum precision. Dithering the
re-sampled PSF over the re-sampled pixel grid gives a spectrum of
non-Gaussian intensity error at each wavelength. We find that these
fluctuations are highly correlated at different wavelengths, the
magnitude and nature of the correlation depending on where each
wavelength falls relative to the pixel grid.  Since NIRSpec exoplanet
observers will $decorrelate$ these fluctuations, we perform that
decorrelation in our simulations.  Specifically, we take two
wavelengths separated by one spectral resolution element (2 pixels),
and we decorrelate intensity fluctuations at the first wavelength with
respect to intensity fluctuations at the second wavelength. We thus
use two fiducial wavelengths to construct a reservoir of intensity
fluctuations that survive the decorrelation process. We call these
`fundamental' fluctuations, and we draw from this reservoir when
modeling other wavelengths, using a random selection process similar
to our MIRI model. In real observations, observers will of course deal
with this effect over all wavelengths simultaneously. Nevertheless,
our procedure encapsulates the essence of the data analysis process
that we envision, and it defines a core of fundamental fluctuations
will not be correctable in exoplanet observations.  These fluctuations
(not illustrated) have a magnitude and non-Gaussian distribution
similar to the MIRI photometry fluctuations (Figure~7).

\section{TESS Planets as Observed by JWST}

We couple all of the specific parameters for each simulated TESS
planet (impact parameter, temperature, orbital period, etc.) to the
JWST instrument noise models, to predict the number of TESS planets of
different types that JWST can characterize. As noted in Sec.~5, we
base the SNR values in this Section on observing all possible
transits/eclipses of each planet, within JWST viewing constraints.

\subsection{MIRI Filter Photometry}

Our values for the SNR of MIRI eclipse characterization are obtained
by synthesizing time series data, as described above, for every
observable eclipse of every TESS planet.  These time series data
contain all of the random error (source photon noise, thermal
background, etc.) and systematic non-Gaussian errors as described in
Sec. 5.1.  We solve for the depth of each eclipse, and we use the
scatter in those derived eclipse depths to define the SNR for one
eclipse.  The aggregate SNR for all eclipses is the single eclipse SNR
times the square-root of the number of eclipses that are observed.

Although we have concentrated our MIRI simulations on secondary
eclipses, we point out the significant potential for
`around-the-orbit' observations \citep{knutson07} applied to
superEarths.  A signal of small, or zero, amplitude in such
observations can be used to establish that the planet has an
atmosphere \citep{nutzcharb, seagerdeming}, a fundamental inference.  Note also
that the sensitivity for around-the-orbit observations can in
principle be higher than for an eclipse, because the full orbit affords
more integration time.

In our eclipse simulations, we represent the superEarths' thermal
emission using the model having intermediate content of free hydrogen
by \citet{miller-ricci}. For exoNeptunes, we use a model by
Miller-Ricci for GJ\,436b. Temperature, rather than composition, has
the dominant effect on MIRI filter photometry, hence we scale the
thermal emission from each model as the temperature varies from planet
to planet. This scaling adopts a blackbody continuum appropriate to
the modeled temperature of each planet, but it does not change the
fractional absorption.  The latter approximation is particularly
appropriate for MIRI, that would focus on the $15\,\mu$m band of
CO$_2$.  The fractional absorption of this band is minimally sensitive
to temperature because it is saturated and arises from the ground
state. For the parent stars, we integrate Phoenix model atmosphere
spectra over the filter bands to simulate the stellar signals.

Figure~9 shows a model spectrum of a superEarth in the $5-$ to
$20\,\mu$m spectral region, from \citet{miller-ricci}, and marks the
bandpasses of MIRI filters at $11.3-$ and $15\,\mu$m. Figure~10 shows
the results of coupling the TESS simulations to our MIRI filter
photometry noise model, for superEarths and exo-Neptunes.  For
superEarths we plot the SNR on the {\it difference} in contrast
between the $11.3-$ and $15\,\mu$m bands, i.e., SNR on the absorption
detection.  For the exo-Neptunes, we plot the SNR in the $15\,\mu$m
band alone, but the SNR for Neptunes is sufficiently high to
contemplate much higher spectral resolution results as well.  SNR will
scale down from Figure~10 as the square-root of the number of
eclipses. Neptune-sized planets will generally not require photometry
of large numbers of eclipses in order to achieve good SNR. Figure~11
shows an example of simulated MIRI 15\,$\mu$m photometry for 10
eclipses of a warm exoNeptune.

Eight habitable superEarths appear on Figure~10, but with lower SNR than for
NIRSpec. This lower SNR arises in part from the background-limited
nature of MIRI observations.  The planet signal scales as the inverse
square of the distance. However, as distance increases the noise
approaches a constant level determined by the thermal background, hence 
$SNR \sim d^{-2}$.  We will see below that NIRSpec characterization of
habitable superEarths attains higher SNR on average, because the
NIRSpec SNR is predominately source-photon-limited and hence $SNR \sim
d^{-1}$, not $\sim\,d^{-2}$.  However, we expect the nearest habitable
TESS superEarth to orbit a nearby ($d \lesssim 10$\,pc) late-M dwarf, and
MIRI will readily detect its secondary eclipse \citep{charbdem}. We
verified that the \citet{charbdem} projections, as updated for their
two cases by \citet{vulcan}, remain approximately consistent with our
current MIRI noise model.

Many hotter superEarths are present on Figure~10 (filled circles), and
their possible CO$_2$ absorption can be characterized by JWST to SNR
$\sim 10$ or greater. Specifically, there are 446 superEarths above
$SNR =10$ on Figure~10.  Although our simulation adopts exactly one
planet per star, only $40$\% of our planets are superEarths, close to
the superEarth frequency claimed by \citet{mayor}. Adopting the
\citet{mayor} $30$\% frequency, TESS will discover $\sim 330$ hot
superEarths whose CO$_2$ absorption can be measured by MIRI filter
photometry.  Above the radius of superEarths ($> 3R_\oplus$ as defined
by TESS), TESS will find many Neptune-size planets (up to $5R_\oplus$)
whose eclipses can be measured by MIRI to high SNR (Figure~11.)

\subsection{NIRSpec Spectroscopy}

We generate synthetic spectroscopy for a given synthetic planet by
including the photon noise from the star, which can vary significantly
with wavelength at the $R=1000$ resolving power of NIRSpec, due to
absorption features in the stellar spectra, especially for the
M-dwarfs.  We use Phoenix model spectra to represent the stars, and we
adopt planet models from \citet{miller-ricci}, as described in the
MIRI case.  Since we are here dealing with transmission spectra, we
scale the magnitude of the absorption from the fiducial planet
model(s) in proportion to the calculated scale height of each planets
atmosphere, and the circumference of its atmospheric annulus.  The
spectra of the fiducial models have been calculated in full detail on
a line-by-line basis by \citet{miller-ricci}.  We preserve this
spectrum shape, and scale the depth of the transmission spectrum in
proportion to the projected area of the atmosphere.  Note that the
geometry of this problem causes an atmosphere to be opaque over a
greater span in scale heights for larger planets.  We include a
radius-dependent factor to account for this effect. Our synthetic
time-series spectroscopy includes fundamental fluctuations due to the
uncorrectable portion of the intra-pixel effect, as described above.
For this purpose we use a large collection of synthetic 10-sec
exposures, and we draw a series of this nonGaussian noise starting at
a random time for each transit of each planet.

Computation of the SNR for each wavelength in the transit spectrum
proceeds by constructing multiple transit curves at each wavelength,
including synthetic photon noise and intra-pixel fluctuations, and
evaluating the SNR of the ensemble of transits, as described above for
MIRI eclipses. The number of observable transits is taken to equal the
number that can be observed for each planet by JWST during its 5-year
mission, under the constraint of JWST's field-of-regard.  Having
evaluated the SNR at each wavelength, we also calculate a total SNR
for the entire band, via a quadrature summation of the SNR values at
each wavelength: $SNR_{band}=\sqrt(\sum snr^2_{i})$, where the sum is
over the number of wavelength channels in the spectrum, and $snr_i$ is
the SNR at a single wavelength $i$, averaged over all observed
transits. When including the transmission signal in this calculation,
we of course use only the variation in transmitted intensity {\it vs.}
wavelength, not the total absorption due to the entire planet.
Figure~12 shows the results of this calculation ($SNR_{band}$) for water absorption
near $2\,\mu$m, and Figure~13 shows the corresponding result for
detection of CO$_2$ absorption at $4.3\,\mu$m.

Figure~14 shows an example of $2\,\mu$m water absorption in both a hot
and habitable superEarth, and Figure~15 shows examples of CO$_2$
absorption in similar planets.

\section{Discussion and Conclusions}

\subsection{Our Results}

Using our default distribution that orbital sizes are distributed
uniformly in $\log(a^{'})$ (Sec.~3.1) places about $15$\% of our
Monte-Carlo planets within the HZ of their star.  Thus, from Figure~1
(interpolating between the 0.1 and 0.3 curve), we project that the
nearest transiting habitable planet will lie about 10 parsecs distant
from our Sun. The nearest transiting habitable planet produced by our
simulation lies at 9.5 parsecs from our Sun.  Hence our simulation is
consistent with the calculation presented in Sec.~2. Moreover, as we
show below, our sensitivity calculations are also internally
consistent because our zeroth order results (Sec.~4) agree well with
our more exhaustive calculations (Secs.~5 \& 6).  We therefore discuss
the prospects for habitable superEarth discovery and characterization
on the basis of our consistent, end-to-end simulation.

The above statements concerning nearby habitable planets are not
restricted to superEarths, because our orbit distributions and radius
distributions are independent.  The architecture of our own solar
system suggests that small planets should be associated with closer
orbits, but we impose no such condition in our simulation.  However,
small planets outnumber large ones in our distributions, and the
nearest transiting habitable planets we generate turn out to be
superEarths, not Neptunes or Jupiters.  Moreover, TESS finds these
nearby habitable worlds very efficiently.  Repeating our simulation
500 times, and considering only the superEarths, we tabulated the
number of times that TESS finds at least one habitable superEarth
closer than 35 pc.  We adopted 0.3 superEarths per star \citep{mayor},
with our default distribution in orbital radius.  The \citet{mayor}
estimate is a frequency: $30\%$ of stars have at least one planet in
the range up to Neptune-sized.  The fact that \citet{mayor} include
Neptunes tends to make our 0.3 superEarths-per-star value too high,
but the fact that many stars will have multiple planets works in the
opposite direction.  Thus we believe that our 0.3 superEarths-per-star
value is reasonable, and it produces 0.047 superEarths per star in the
habitable zone. With this abundance, TESS finds at least one
transiting habitable superEarth closer than 35 parsecs in 495 out of
500 simulations, a $99\%$ probability.


Nearby habitable transiting superEarths can be characterized to a
significant degree by JWST.  The level of significance for this
characterization is a function of astrophysical uncertainties, not
primarily technological ones.  First, consider the situation for
near-IR water absorption measured by transmission spectroscopy at
transit.  Our zeroth order calculation (Figure~4) indicates that water
absorption could be measured in 5 nearby habitable superEarths to SNR
equal or exceeding 10, versus 8 from Figure~12.  A key difference in
these figures is that Figure~4 adopts a 200-hour program, whereas
Figure~12 is based on observing all transits available within the JWST
mission lifetime.  But habitable planets tend to have longer orbit
periods (compared to `hot superEarths'), resulting in longer transits
that occur less frequently. Inspecting the number of habitable
superEarth transits available to JWST per system, we find a natural
dividing point near 60 transits. Five of the eight Figure~12 habitable
superEarths exhibit 60 or fewer transits, and the total in-transit
time for each of them is less than 200 hours. These five habitable
superEarths would be suitable to observe using a large JWST program.
An example of synthetic JWST data for water absorption in a habitable
superEarth with relatively low aggregate SNR ($SNR_{band}=16$) is shown in the
lower panel of Figure~14.  SuperEarths represent about $40$\% of the
planets in our simulation, close to the $30$\% frequency claimed by
\citet{mayor}.  (Hence, our simulations can be scaled to the
\citet{mayor} frequency by multiplying the number of superEarths by
$0.75$.) If their frequency is lower than \citet{mayor} claim, then
the number available for JWST characterization will be reduced in
proportion.

Another source of astrophysical uncertainty regarding JWST
characterization is the nature of the superEarth atmosphere.  We have
used the intermediate case of \citet{miller-ricci}, where the
atmosphere is mildly reducing and inflated in height by the presence
of residual hydrogen (either primordial, or outgassed).  In the
absence of this hydrogen, the mean molecular weight increases, the
scale height decreases, and the water abundance also decreases.  In
that case, the SNR for transmission spectroscopy drops by a factor of
$\sim 3$, but JWST characterization remains possible for the closest
superEarths.  Specifically, five habitable superEarths would remain
above $SNR_{band}=10$, of which about three could be observed in 60 or fewer
transits.  If the frequency of superEarths is as low as $0.1$ per
star, and their atmospheres are hydrogen-depleted, then the number
capable of being characterized in water absorption by JWST is expected
to be {\it one}. On the other hand, if the frequency estimate of
\citet{mayor} is correct, and also their atmospheres are mildly
reducing ($10$\% hydrogen), then the expectation for the number that
JWST can characterize via water absorption is {\it five}.

The situation for detection of CO$_2$ in absorption at $4.3$\,$\mu$m
during transit is similar to that of water.  Figure~13 indicates seven
habitable superEarths which could be observed to $SNR \geqq 10$, of
which four can be observed in less than 60 transits. Synthetic data
for one example is shown in the lower panel of Figure~15.  This
habitable superEarth has $SNR_{band}=28$, obtained by observing 58
transits. If the atmospheres of these worlds are hydrogen-depleted,
then the SNR drops by a factor of two, and the number remaining above
$SNR=10$ in 60 transits or less drops to two.  Factoring in possible
variation in superEarth frequency (as above), then our expectation for
the number capable of being characterized in CO$_2$ absorption by JWST
to $SNR=10$ ranges from {\it one} to {\it five}.

The lower panel of Figure~4 indicates 4 habitable superEarths detected
in thermal continuum radiation at secondary eclipse to $SNR \geqq
10$. Simulation of CO$_2$ absorption observations using MIRI
photometry at $15$\,$\mu$m (Figure~10) indicate one superEarth having
$SNR \geqq 10$, with an additional one at $SNR=8$. These smaller
numbers are consistent with the more demanding measurement of
absorption {\it vs.} a simple continuum brightness temperature
measurement.  An additional 3 are present on Figure~10 with $SNR \gtrapprox 4$,
sufficient to establish at least the presence of CO$_2$ absorption.
Unlike the case of transit spectroscopy, this secondary eclipse
technique is not sensitive to the scale height of the atmosphere, only
to the total column density of absorbing molecules.  Therefore we need
only consider the frequency of superEarths.  On this basis we expect
the number capable of being characterized via a thermal continuum
temperature measurement, and identification of CO$_2$ absorption is
from {\it one} to {\it four}.

Although our projections as quoted above are from a single Monte-Carlo
simulation of the TESS yield, we have re-run our simulations multiple
times to verify the stability of the results in a statistical sense.
We emphasize that JWST characterization of habitable worlds will
require a large program, with priority given to the available
transits/eclipses.  Also, difficult choices will be necessary since
each transit is precious but can be observed in only one mode
(instrument, grating setting, etc.) at a time. Although the number of
habitable planets capable of being characterized by JWST will be
small, large numbers of warm- to hot-superEarths and exoNeptunes will
be readily characterized by JWST, and their aggregate properties will
shed considerable light on the nature of icy and rocky planets in the
solar neighborhood.

\subsection{Comparison to Other Results}

Pioneering work by \citet{valenti06} considered the possibility of
JWST/NIRSpec characterization of habitable terrestrial planets that
may transit nearby M-dwarfs, based on models by \citet{ehrenreich}.
They find that a habitable ocean planet orbiting an M3V star at 13
parsecs distance (J-magnitude = 8) can be well characterized using
NIRSpec transmission spectroscopy.  However, they find that water and
CO$_2$ absorption in an Earth-sized planet with a high molecular
weight atmosphere, i.e., a true Earth analog, can be detected only if
the planet orbits an unrealistically bright and nearby M-dwarf (J=5,
lying at 3 parsecs distance).  Our calculations specifically include
realistic details such as the number of transits available within the
JWST field-of-regard, effects of non-central transits, systematic
errors in the instrument, etc.  However, we find that inclusion of
these realistic factors does not greatly degrade the JWST sensitivity,
and we concur with the \citet{valenti06} results.  Specifically, we
note that none of the habitable planets above $SNR=10$ on Figure~12
are true Earth analogs, but this does not diminish their importance. 
All of them have radii exceeding $1.8\,R_\oplus$, and most of them are
ocean planets.

Recently, \citet{kaltenegger} have calculated the detectability of
Earth's transmission spectrum, illuminated by lower main sequence
stars at 10 parsecs. They find that the transmission spectrum of our
Earth has $SNR < 1$ for every molecular band in a single transit.
However, many of the molecular features they tabulate would be
observed to $SNR > 10$ in a 200-hour program (their Table~3).

Regardless of the situation for a true Earth analog, results for Earth
cannot be easily extrapolated to superEarths.  The atmospheric scale
height is proportional to $1/g$, where $g$ is surface gravity.  If we
make the reasonable assumption that mass density is constant as radius
varies, then $g \sim M/R_p^2 \sim R_p$. The projection of an
atmospheric annulus of a given scale height is then proportional to
$R_p/g$, and thus absorption seems independent of $R_p$.  However,
superEarths include ocean planets having lower mass densities
\citep{seager07, fortney}, which invalidates the assumption of
constant density.  Also, slant-path absorption does not scale strictly
with $H$ as $R_p$ increases.  In the slant-path geometry, for a given
line opacity (cm$^2$/g), the atmosphere is opaque over more scale
heights as $R_p$ increases.  This modest but significant factor is
included in our calculations.

We reiterate our conclusion that, depending on the frequency of
occurrence of superEarths and the nature of their atmospheres, JWST
will be able to measure the temperature, and detect molecular
absorption bands, in one to four habitable TESS superEarths.

\acknowledgments

T. Greene and M. Clampin gratefully acknowledge support from the
JWST Project. We thank J.~Valenti for sending us his exoPTF White
Paper, and Lisa Kaltenegger for an advance copy of her ApJ paper.  We are
grateful to Tilak Hewagama for a clarifying discussion on
simulation of JWST pointing jitter, and to the referee for helpful comments that
significantly improved the manuscript.

\clearpage



\begin{figure}
\epsscale{.60}
\plotone{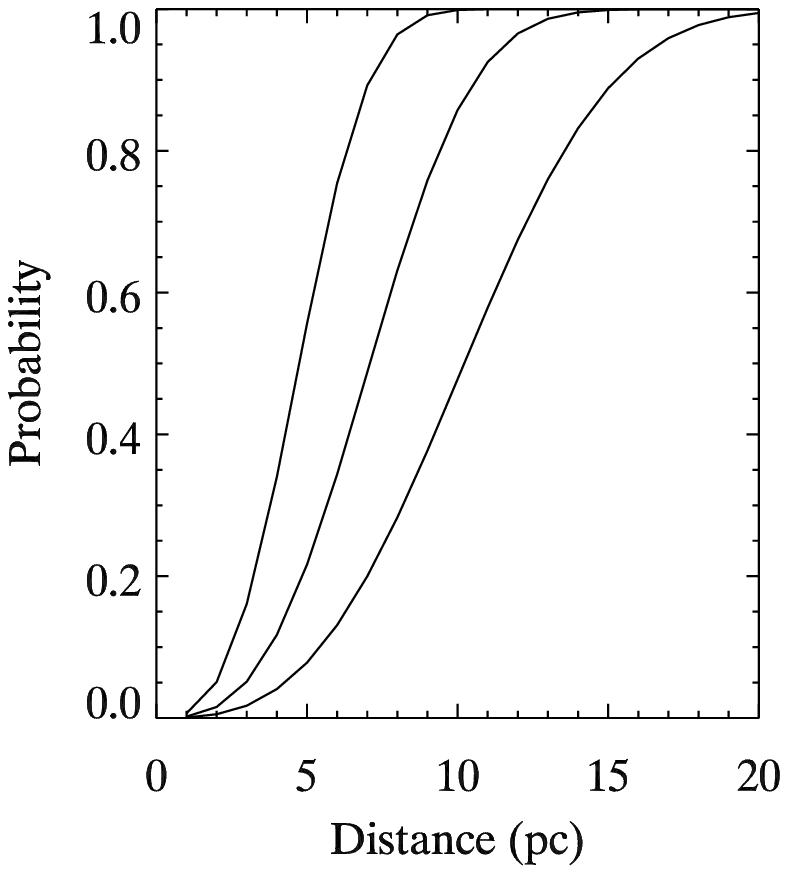}
\vspace{0.5in}
\caption{Probability of a habitable transiting planet lying within a
sphere of a given distance (radius) centered on the Sun, for planet
frequencies of (left to right) $1.0$, $0.3$, and $0.1$ per star,
adopting the condition that all planets lie within the HZ (see text).}
\end{figure}

\clearpage

\begin{figure}
\epsscale{.60}
\plotone{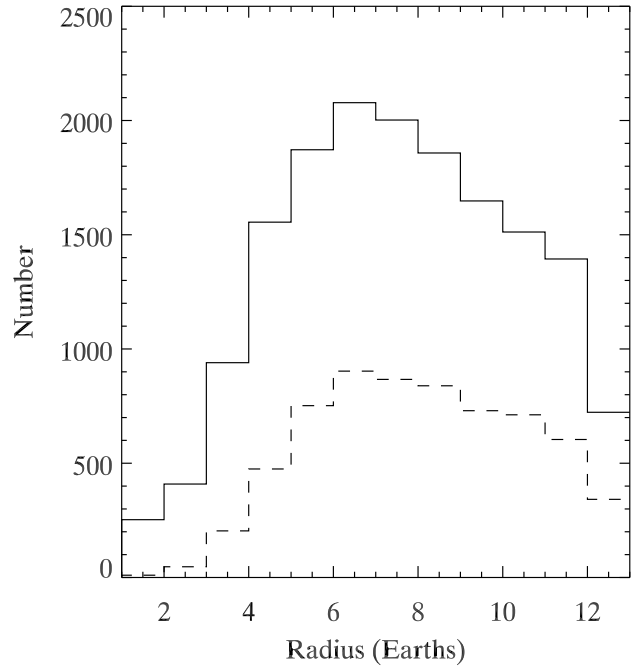}
\vspace{0.5in}
\caption{Number of planets detected by TESS in our simulations {\it
vs.} radius, using two different distributions of planet orbital
distance.  The solid line is our default distribution, that places planets
with uniform probability in $\log a$, scaled by the stellar
luminosity (see text). The dashed line places planets with uniform
probability in $a$, between 0.05 and 1.0 AU.}
\end{figure}

\clearpage

\begin{figure}
\epsscale{.50}
\plotone{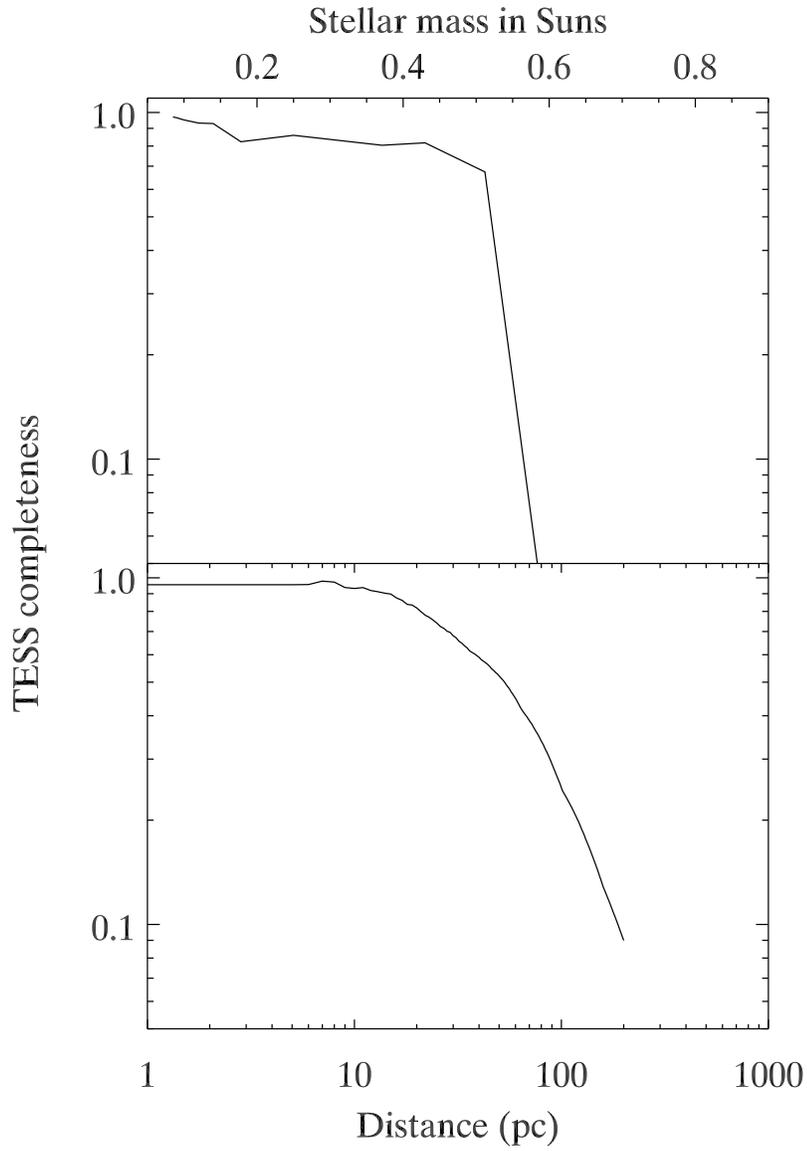}
\vspace{0.4in}
\caption{{\it Upper panel:} Completeness of the TESS survey for the
detection of transiting superEarths ($1-3R\oplus$) orbiting main
sequence stars of different masses, in a volume out to 35 parsecs
distance.  {\it Lower panel:} Completeness of the TESS survey for the
detection of transiting superEarths orbiting in the habitable zone of
stars of all spectral types, versus distance.}
\end{figure}

\clearpage

\begin{figure}
\epsscale{.50}
\plotone{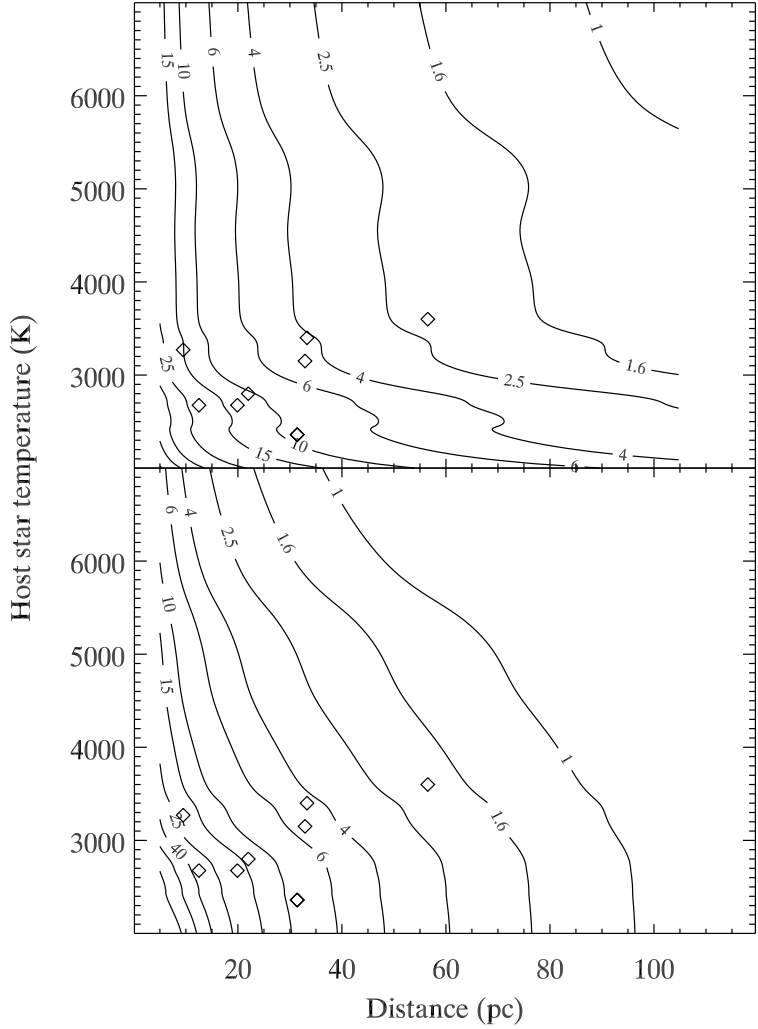}
\vspace{0.3in}
\caption{{\it Upper panel:} Contours of signal-to-noise ratio (SNR) for an
intermediate composition $10\,M_\oplus$ habitable superEarth
($R=2.3R_\oplus$, $T=323$K) observed during transit by a large
aperture space telescope (area $25\,m^2$), observing water absorption
at 2\,$\mu$m, using a spectral resolving power of 1000, and summing
the SNR over the 160 most optically thick wavelengths.  {\it Lower
panel:} SNR contours for the secondary eclipse of the same habitable
superEarth, observed in thermal continuum radiation using a
large-aperture cryogenic space telescope at 15\,$\mu$m, with a
3\,$\mu$m optical bandwidth (FWHM).  Both panels assume the average of
multiple transits or eclipses observed during a 200-hour program. The
overplotted points show the TESS-discovered planets from our
simulation, choosing all planets having radii $R=2.3\pm
1.0\,R_{\oplus}$, and $T=323\pm50\,K$.}
\end{figure}

\clearpage

\begin{figure}
\epsscale{.50}
\plotone{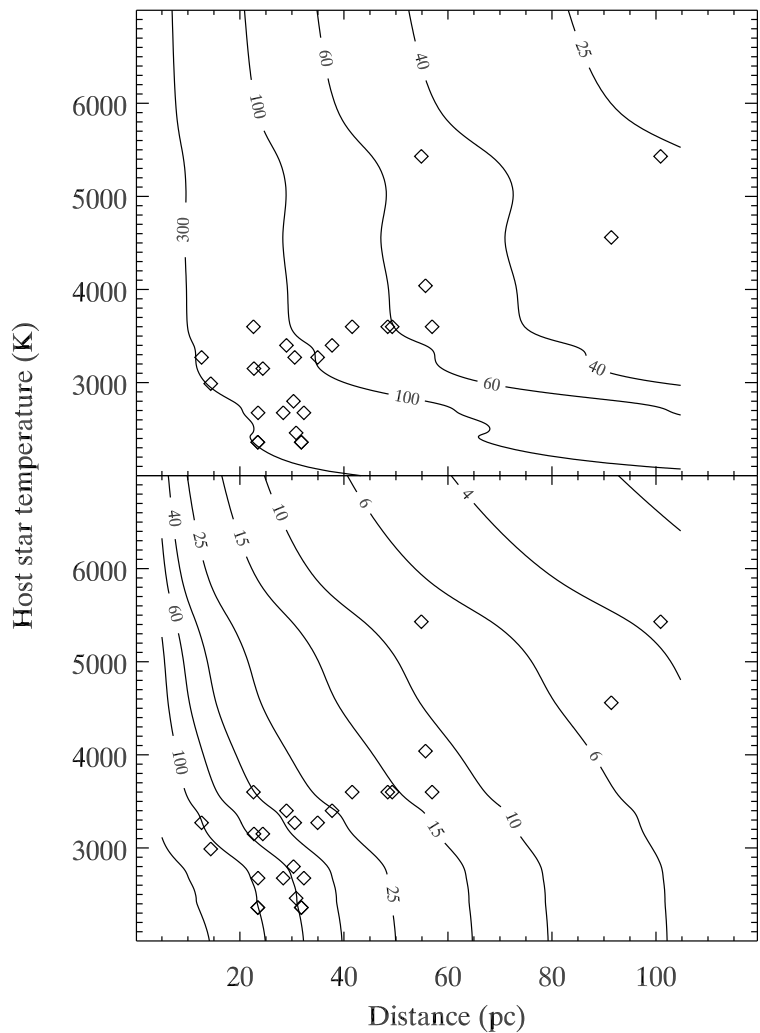}
\vspace{0.3in}
\caption{Like Figure~3, but using a planet of $20\,M_\oplus$,
$R=2.7R_\oplus$, hydrogen-helium-dominated atmospheric composition, and
$T=500$K. The overplotted points are all TESS-discovered planets from
our simulation, having $R=2.7\pm 1.0\,R_{\oplus}$, and
$T=500\pm50\,K$.}
\end{figure}

\clearpage

\begin{figure}
\epsscale{.80}
\plotone{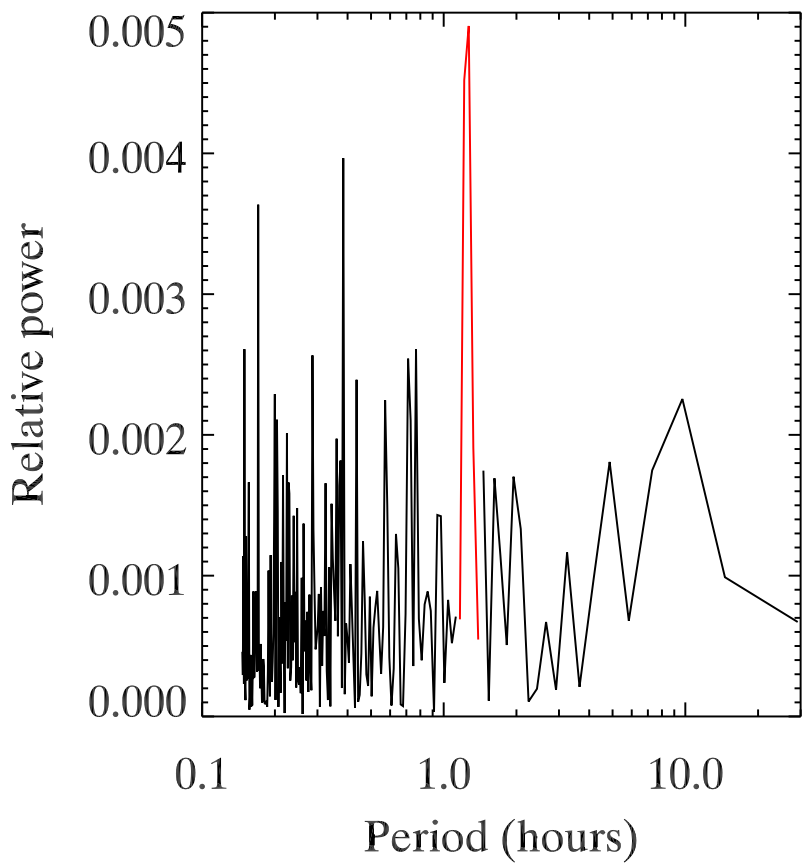}
\caption{Fourier power spectrum of a 33-hour sequence of {\it Spitzer}
aperture photometry \citep{seagerdeming}, showing a peak (in red) due
to a pointing oscillation in the telescope, with a period slightly
longer than 1 hour.  We use these data to estimate the pixel-to-pixel
flat-fielding error for this mode of {\it Spitzer} observations, and
then project this effect to JWST (see text).}
\end{figure}

\clearpage

\begin{figure}
\epsscale{.60}
\plotone{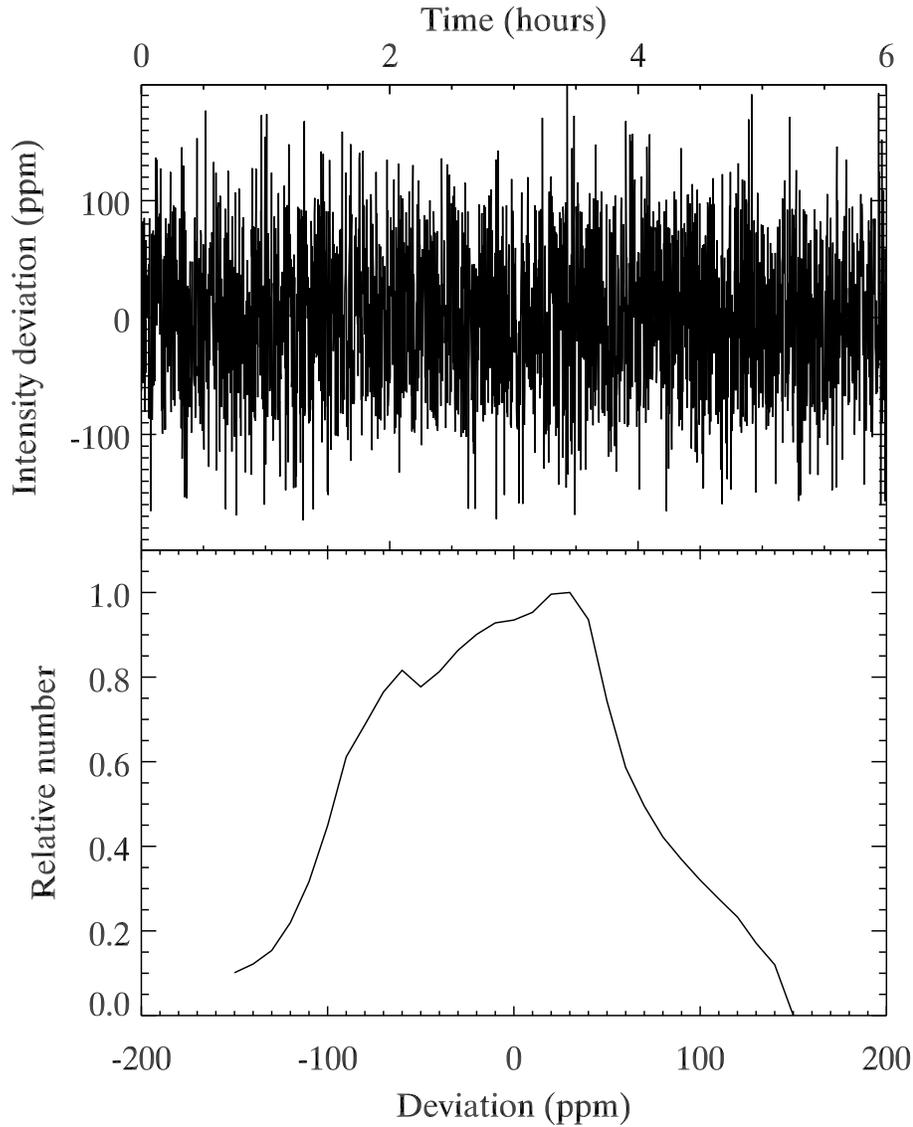}
\vspace{1.0in}
\caption{{\it Upper panel:} Intensity deviation in parts per million
(ppm) for JWST/MIRI photometry at $15\,\mu$m, from our MIRI noise
model.  These deviations are due to telescope pointing fluctuations
and pixel-to-pixel errors in flat-fielding, based on {\it Spitzer}
experience.  {\it Lower panel:} Histogram of deviations from the upper
panel, showing the non-Gaussian character of this noise source.}
\end{figure}

\clearpage

\begin{figure}
\epsscale{.70}
\plotone{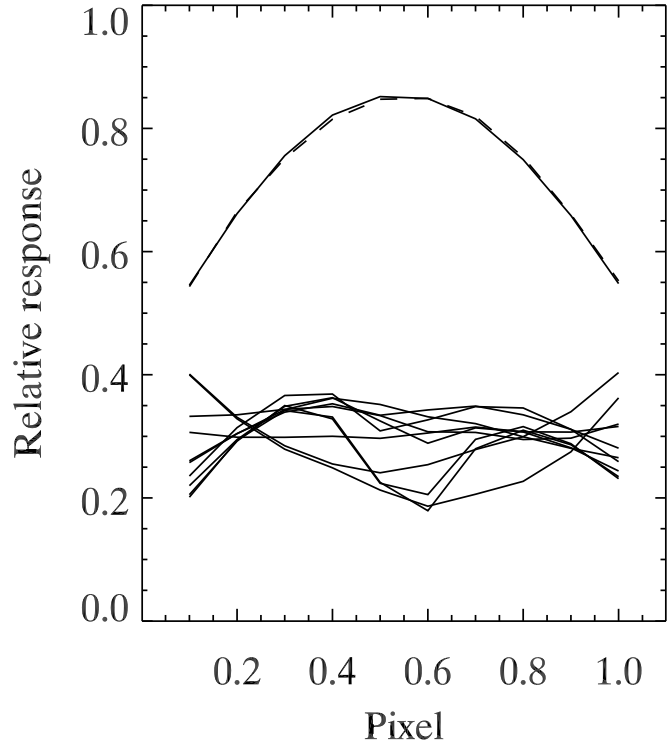}
\vspace{1.0in} \caption{Intrapixel sensitivity variation for a
representative NIRSpec detector pixel, from engineering measurements
of the flight detector.  The top traces show the average variation in
the dispersion direction (solid line), and the spatial direction
(dashed line).  The lower traces divide the pixel into ten strips 
parallel to the spectral dispersion, and they show the difference from
a parabolic fit of response versus distance from pixel center.  The
differences have been amplified by a factor of four, and offset by
0.3, for clarity of presentation.}
\end{figure}

\clearpage

\begin{figure}
\epsscale{.60}
\plotone{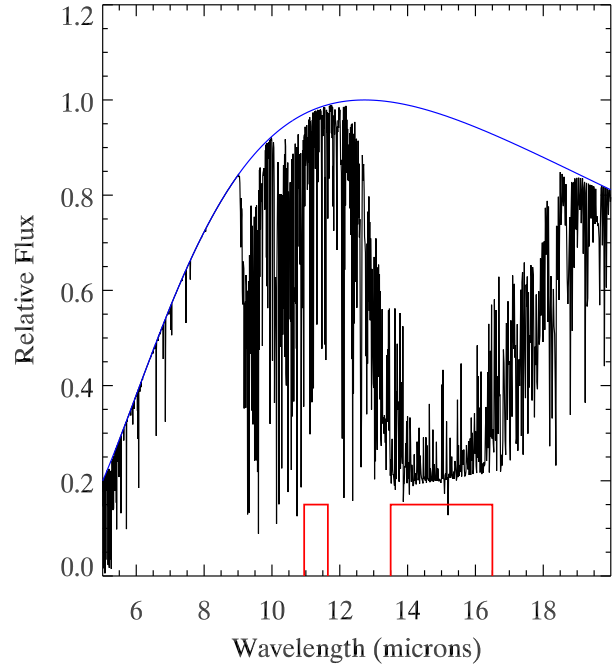}
\vspace{1.0in} \caption{Relative flux versus wavelength for a
super-Earth model \citep{miller-ricci}.  The blue line shows the
blackbody continuum, and the red brackets denote the bandpasses of
MIRI filters at $11.3-$ and $15\,\mu$m. }
\end{figure}

\clearpage

\begin{figure}
\epsscale{.70} \plotone{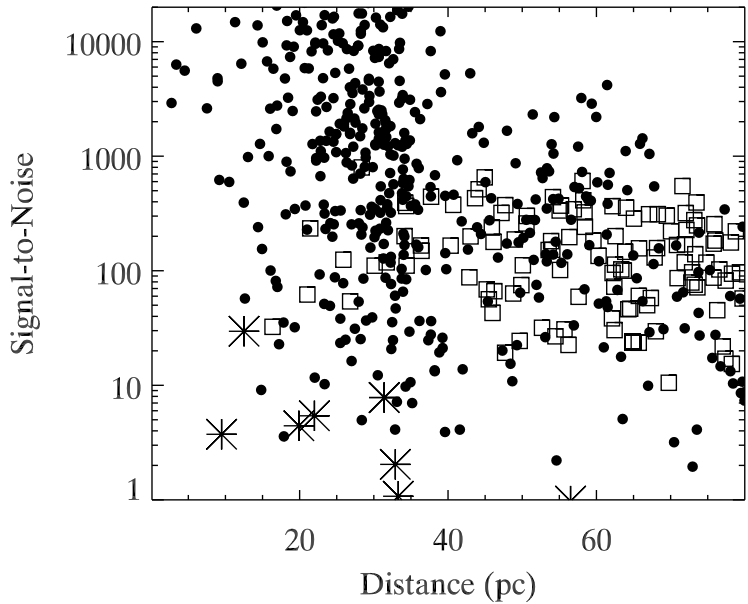}
\vspace{1.0in}
\caption{Signal-to-noise for TESS planets, measured by MIRI filter
photometry at $15\,\mu$m (the most favorable wavelength), versus
distance in parsecs.  The SNR for superEarths applies to detection of
the CO$_2$ absorption at $15\,\mu$m relative to $11.3\,\mu$m (see
text). Stars are habitable superEarths, and filled points are
superEarths having equilibrium temperatures above 373\,Kelvins.  Open
squares are planets with radii between 3 and 5 Earth radii (Neptunes),
at all temperatures, and their SNR is for detection of continuum
radiation at $15\,\mu$m. Points at the highest SNR tend to be hot
planets in short-period orbits, lying at high ecliptic latitude where
JWST has access to a very large number of eclipses.}
\end{figure}

\clearpage

\begin{figure}
\epsscale{.70} \plotone{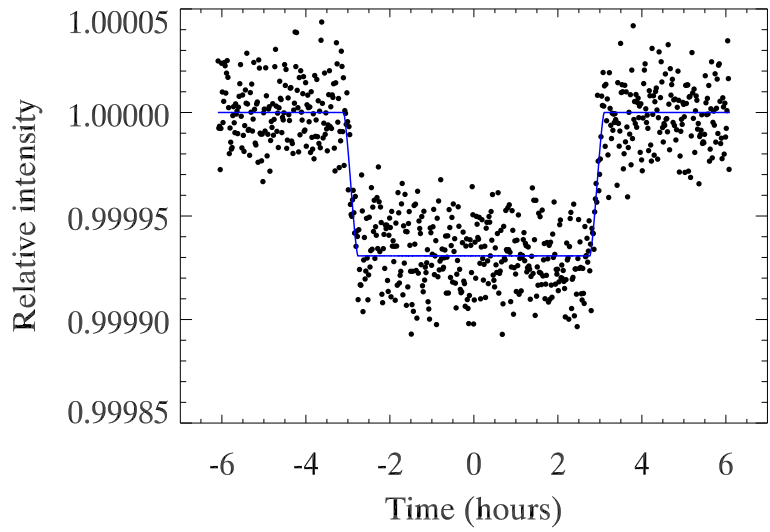}
\vspace{1.0in}
\caption{Example of synthetic JWST/MIRI secondary eclipse photometry
for a warm ($T=500K$) exoNeptune, averaging 10 eclipses (120 hours
total observing) as observed by JWST/MIRI at $15\,\mu$m. The synthetic
observations were binned to 1-minute time resolution. This planet has
$R=4\,R_\oplus$, and orbits at $a=0.2$\,AU from a K2V star.}
\end{figure}

\clearpage

\begin{figure}
\epsscale{.70} \plotone{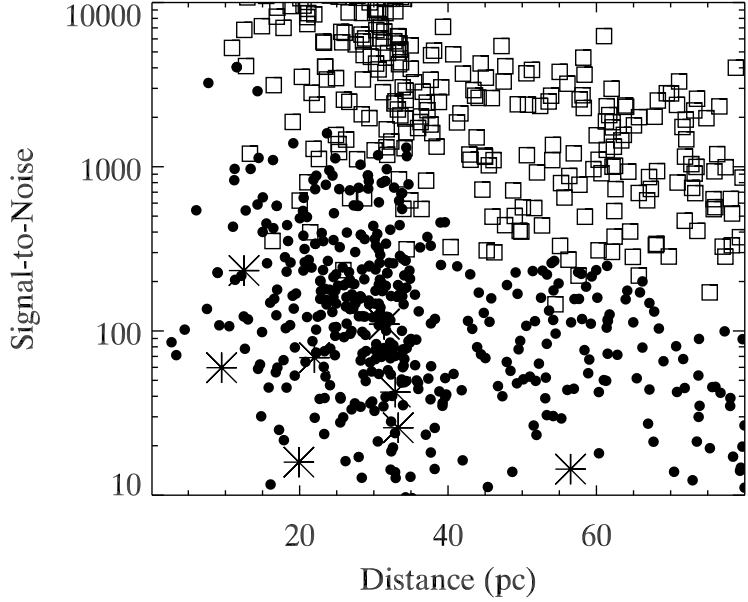}
\vspace{1.0in}
\caption{Signal-to-noise ratio for water absorption in TESS planets,
measured by NIRSpec $R=1000$ spectroscopy at $2\,\mu$m, versus
distance in parsecs. The SNR is for the integral over wavelength from
$1.7-$ to $3.0\,\mu$m (see text). Solid circles are superEarths having
equilibrium temperatures above 373\,Kelvins. Stars are superEarths at
habitable temperatures. Open squares are planets with radii between 3
and 5 Earth radii (Neptunes), at all temperatures. Unlike Figs.~3 \&
4, we here invoke the condition that JWST observes all possible
transits of a given system during its 5-year mission. Points at
the highest SNR tend to be hot planets in short-period orbits, lying
at high ecliptic latitude where JWST has access to a very large number
of transits.}
\end{figure}

\begin{figure}
\epsscale{.70} \plotone{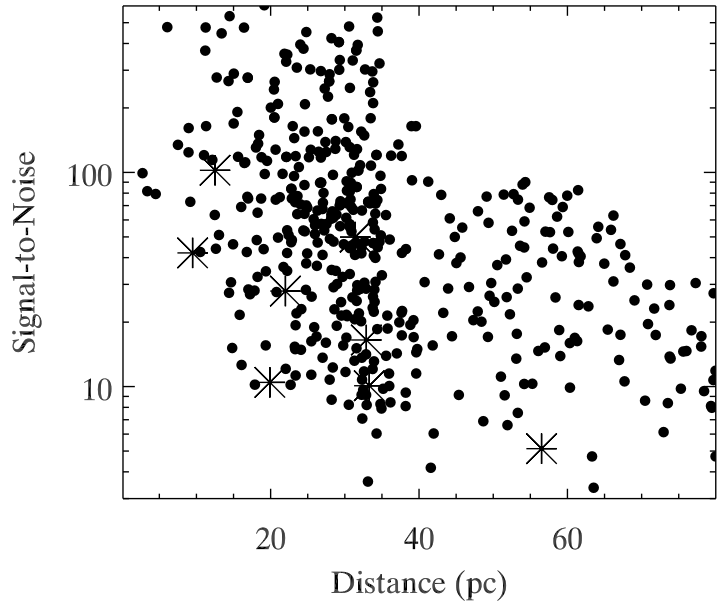}
\vspace{1.0in}
\caption{Signal-to-noise ratio for TESS planets in the $4.3\,\mu$m
CO$_2$ band, measured by NIRSpec $R=1000$ spectroscopy, versus
distance in parsecs. The SNR is for the integral over wavelength from
$4.0-$ to $4.6\,\mu$m, i.e. the total SNR for the band. Solid circles
are superEarths having equilibrium temperatures above 373\,Kelvins,
and stars are superEarths at habitable temperatures. Points at
the highest SNR tend to be hot planets in short-period orbits, lying
at high ecliptic latitude where JWST has access to a very large number
of transits.}
\end{figure}

\begin{figure}
\epsscale{.50} \plotone{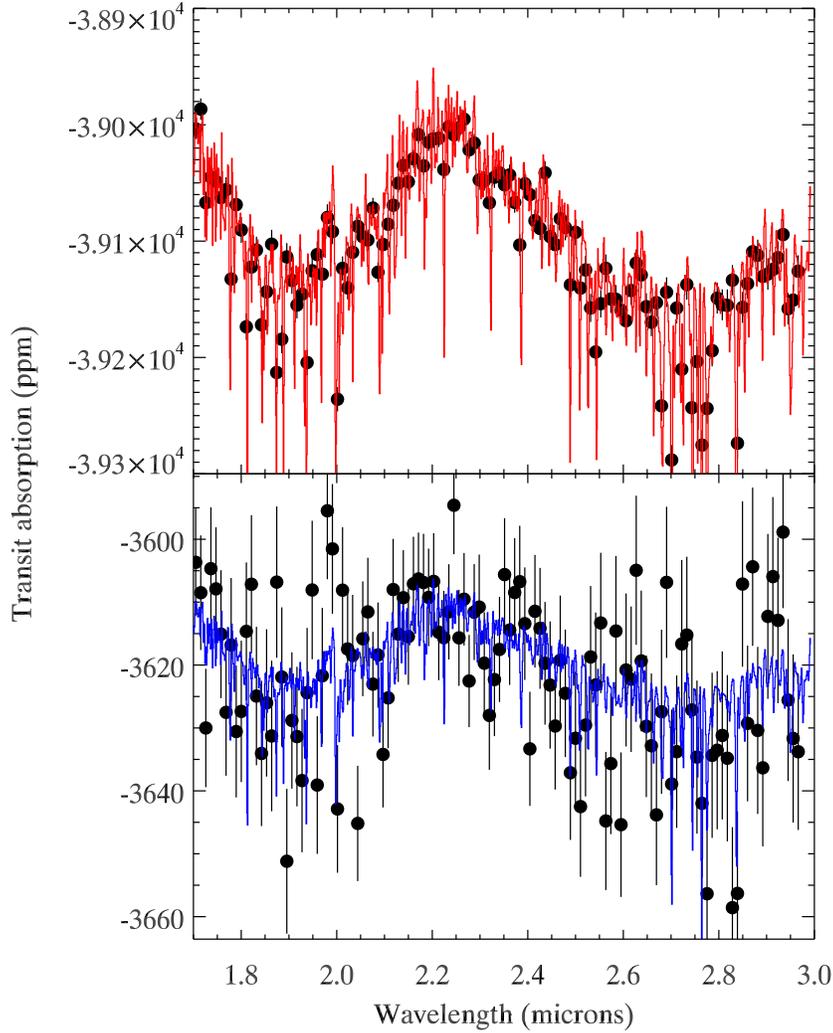}
\vspace{0.5in}
\caption{{\it Upper panel:} Points are synthetic NIRSpec observations
of water absorption near $2\,\mu$m, in a hot ($T=506$K) superEarth
having $R=2.1R_{\oplus}$, at a distance of $32$ parsecs.  The red line
is the modeled spectrum, and the synthetic data have been binned in
wavelength by a factor of 10, to a spectral resolving power $R=100$,
for clarity of presentation.  The SNR for the aggregate detection of
water absorption in this example is $SNR=163$, for 301 hours.  {\it
Lower panel:} synthetic NIRSpec observations of water absorption in a
habitable superEarth having $T=302$K and $R=1.8R_{\oplus}$.  The
aggregate SNR for this detection is $SNR=16$ for 122 hours, and the
distance to this M-dwarf planetary system is $d=20$\,parsecs.}
\end{figure}

\begin{figure}
\epsscale{.50} \plotone{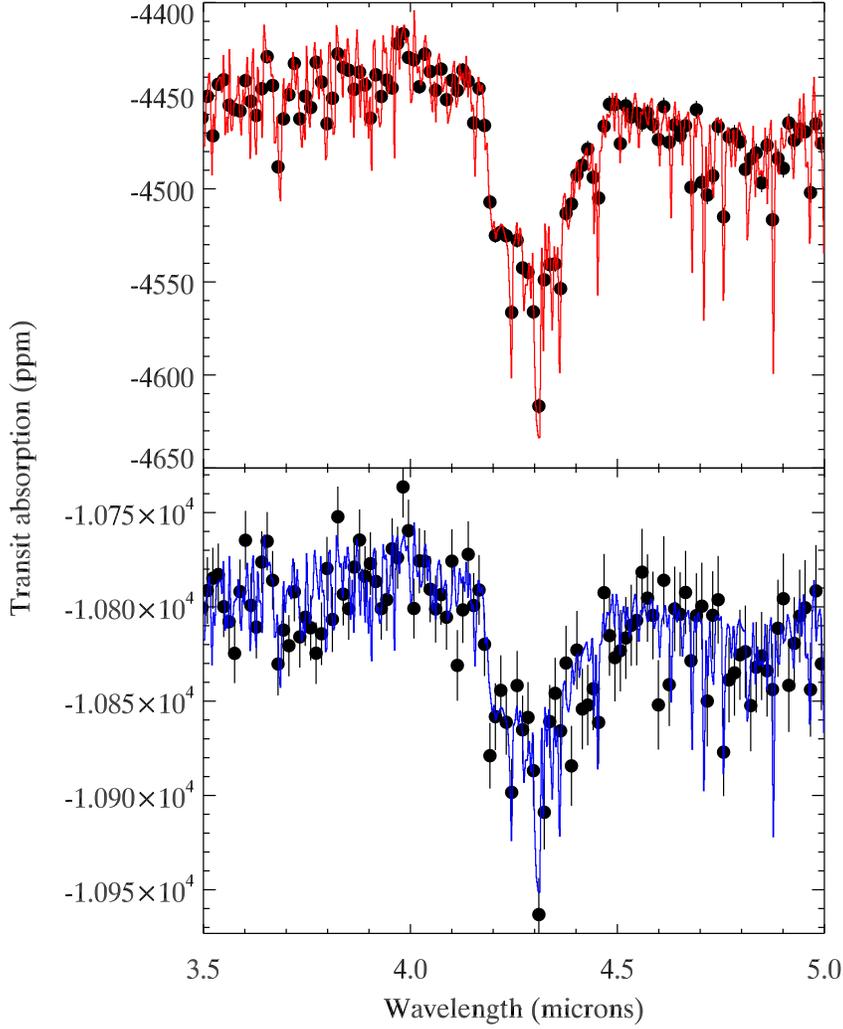}
\vspace{0.5in}
\caption{{\it Upper panel:} synthetic NIRSpec observations (points) of
carbon dioxide absorption near $4.3\,\mu$m, in a hot ($T=797$K)
superEarth having $R=2.2R_{\oplus}$, at a distance of $18$ parsecs,
with the model overlaid (red line).  The synthetic data have been
binned in wavelength by a factor of 10, to a spectral resolving power
$R=100$, for clarity of presentation.  The SNR for the aggregate
detection of water absorption in this example is $SNR=150$, observing
for 480 hours in-transit.  (The SNR for 20-hours in-transit would scale
down to $31$.) {\it Lower panel:} synthetic NIRSpec observations
(points) of carbon dioxide absorption in a habitable superEarth having
$T=308$K and $R=2.3R_{\oplus}$, with the model specturm overlaid (blue
line).  The aggregate SNR for this detection is $SNR=28$ for 85 hours
of in-transit observing, and the distance to this planetary system is
$d=22$\,parsecs.}
\end{figure}

\end{document}